\documentclass[10pt,a4paper]{article}
\usepackage[a4paper,left=2.5cm,right=2.5cm,top=2.5cm,bottom=2.5cm]{geometry}
\usepackage[super,compress]{cite}
\usepackage{graphicx}
\usepackage{authblk}
\begin{document}

\title{Tachyon constant-roll inflation in Randall-Sundrum II cosmology}

\author[1]{Marko Stojanovic\thanks{marko.stojanovic@pmf.edu.rs}}
\author[2]{Neven Bili\'c\thanks{bilic@irb.hr}}
\author[3]{Dragoljub D. Dimitrijevic\thanks{ddrag@pmf.ni.ac.rs}}
\author[3]{Goran S. Djordjevic\thanks{gorandj@junis.ni.ac.rs}}
\author[3]{Milan Milosevic\thanks{milan.milosevic@pmf.edu.rs}}

\affil[1]{Faculty of Medicine, University of Ni\v s, Serbia}
\affil[2]{Division of Theoretical Physics, Rudjer Bo\v{s}kovi\'{c} Institute, Zagreb, Croatia}
\affil[3]{Department of Physics, University of Ni\v s, Serbia}

\maketitle

\begin{abstract}
We study inflation in a model with constant second slow-roll parameter $\eta$. In this case, the Hubble expansion rate equation has analytical solutions describing four possible, nontrivial inflation scenarios. The evolution of the inflaton governed by a tachyon field is studied in the framework of the standard and Randall-Sundrum II cosmology. The attractor behavior of the solution is briefly demonstrated. Finally, the calculated values of the parameters $n_{\rm s}$ and $r$ are compared with observational data.

\end{abstract}


\section{Introduction}  

The theory of cosmological inflation is the leading theory in describing the early universe. Inflation solves the flatness, horizon, and other problems in standard  cosmology\cite{Guth:1980zm}. Besides, the quantum fluctuations of the field which govern inflation seed the large-scale structure of the universe. 

The physical mechanism that drives inflation is not entirely known and subject to speculation. However, it is widely accepted that at least one scalar field (inflaton) can describe the inflation mechanism. One of the candidates for inflaton is a tachyon field, whose origin is related to the instability of the perturbative vacuum of string theory\cite{Feinstein:2002aj}. 

This paper aims to analyze the constant-roll inflation with the slow-roll parameter $\eta$ being constant. The tachyon field describes the dynamics of inflation in the framework of the braneworld cosmology, based on the second Randall-Sundrum model (RSII)\cite{Randall:1999vf}. This work is motivated by the model introduced in Ref.  \citen{Mohammadi:2020ftb}, where the idea of the constant-roll was introduced for the canonical scalar field in the framework of RSII cosmology. As a novelty, we apply this approach to the RSII cosmology with a tachyon field. It is known that a tachyon field and the corresponding inflatory expansion can be studied within RSII cosmology\cite{Bilic:2017orf, Bilic:2013dda}. Therefore, as a logical step, we extend the study to include constant-roll inflation. In this setup, the second slow-roll parameter $\eta$ is chosen to be constant as it is usually done in the literature\cite{Yi:2017mxs, Mohammadi:2018oku, Nguyen:2021emx}. However, this choice is applicable only to a canonical scalar field.

In contrast, if applied to the tachyon field inflation, the same choice does not lead to a correct analytical solution because the inflation stage never ends. Instead, we propose a model-independent definition of $\eta$ in terms of the Hubble expansion rate and its derivatives. With this choice, the Hubble expansion rate can be found  analytically. In the following, we calculate the inflation parameters and compare the predictions of the model with Planck results\cite{Planck:2018jri}.

This paper is organized as follows. In Section 2, we introduce the slow-roll parameters. In Section 3, we calculate the parameters $n_{\rm s}$ and $r$ for different solutions of the equation for the Hubble rate. Section 4 is devoted to the  constant-roll tachyon inflation in standard and RSII cosmology. Finally, the paper is concluded in Section 5.

\section{Slow-roll parameters}

The slow-roll approximation is a common way to analyze the inflation models. In order to carry out this procedure, the slow-roll parameters have been introduced\cite{Liddle:1994dx}. The Hubble slow-roll parameters are defined as 
\begin{equation}
\epsilon=-\frac{{\dot{H}}}{H^2},
\label{epsilon}
\end{equation} 
\begin{equation}
\eta=-\frac{\ddot{H}}{2H\dot{H}}.
\label{defeta1}
\end{equation}
In the slow-roll regime, the inflaton field is changing according to the slow-roll assumption
\begin{equation}
\epsilon\ll 1\quad \textrm{and}\quad \eta\ll 1.
\end{equation}
The parameter $\eta$ in the model with canonical scalar field $\phi$ takes the form \cite{Yi:2017mxs}
\begin{equation}
\eta=-\frac{\ddot{\phi}}{H\dot{\phi}}.
\label{eta}
\end{equation}
It is helpful to define the slow-roll parameter hierarchically\cite{Steer:2003yu}
\begin{eqnarray}
\varepsilon_{0}&\equiv &H_{*}/H,\label{de0}\\
\varepsilon_{i+1}&\equiv &\frac{d\ln|\varepsilon_{i}|}{dN},\quad i\geq 0.
\label{dei}
\end{eqnarray}
These parameters satisfy the recurrence relation
\begin{equation}
\dot{\varepsilon}_{i}=H\varepsilon_{i}\varepsilon_{i+1}.
\label{parrec}
\end{equation}
Here $H_{*}$ is the Hubble parameter in some chosen time, and $N$ is the number of e-folds in the exponential expansion of the universe, defined as 
\begin{equation}
N=\int_{t_{\rm i}}^{t_{\rm f}} Hdt.
\label{Ndef}
\end{equation}
Here $t_{\rm i}$ and $t_{\rm f}$ denote the times of the beginning and the end of inflation, respectively.
The slow-roll parameters defined by (\ref{de0}) and (\ref{dei}) are independent of the field driving inflation. Hence, we will express the parameter $\eta$ through  $\varepsilon_{i}$. Combining the definitions of the first two slow-roll parameters
\begin{eqnarray}
\varepsilon_{1}&=&-\frac{\dot{H}}{H^2},\label{eq1a}\\
\varepsilon_{2}&=&\frac{\dot{\varepsilon}_{1}}{\varepsilon_{1}H},
\label{eq1b}
\end{eqnarray}  
with (\ref{epsilon}) and (\ref{defeta1}), one finds\cite{Schwarz:2001vv} 
\begin{equation}
\epsilon\equiv \varepsilon_{1},
\end{equation}
\begin{equation}
\eta=\varepsilon_{1}-\frac{1}{2}\varepsilon_{2}.
\label{defeta}
\end{equation}
The expression (\ref{defeta}) is model-independent and allows us to analyze and compare various inflation models.

\section{Constant-roll inflation}

Equations for inflationary dynamics are simplified in the slow-roll approximation, and in some cases can be solved analytically. However, there are cases when this approach cannot be applied. When a potential posses an extremely flat region, or an inflection point, the conditions for slow-roll have been violated, and inflation passes through the ultra-slow-roll regime\cite{Tsamis:2003px}. In this case,  the second slow-roll parameter $\eta=-\ddot{\phi}/(H\dot{\phi})=3$ becomes constant. The assumption that the parameter $\eta$ can have arbitrary constant value corresponds to constant-roll inflation\cite{Martin:2012pe, Motohashi:2014ppa}. The idea of constant parameter $\eta$ during inflation has been generalized to other slow-roll parameters being constant\cite{Gao:2018tdb}.

This paper aims to study inflation with the constant parameter $\eta$. Using expression (\ref{parrec}) one finds  
\begin{equation}
\varepsilon_{2i}=2\varepsilon_1-2\eta,\quad \varepsilon_{2i+1}=2\varepsilon_1,\quad i\geq 1,
\label{parrec1}
\end{equation}
which can be proved by induction. From expression (\ref{defeta1}) one finds a differential equation
\begin{equation}
\ddot{H} +2\eta H\dot{H}=0.
\label{eq00}
\end{equation}
It is interesting to note that (\ref{eq00}) is invariant under scaling 
\begin{equation}
H\rightarrow \ell H,\quad t\rightarrow t/\ell,
\end{equation}
where $\ell$ has the dimension of length. From now on it is understood that the dimensionful quantities are measured in units of $\ell$ to some power. For example, $H$ is measured in units of $\ell^{-1}$ and $t$ in units of $\ell$. 

A straightforward solution to (\ref{eq00}) is provided by
\begin{equation}
\dot{H} = -\eta H^2,
\label{eq04}
\end{equation}
yielding
\begin{equation}
H=\frac{1}{\eta t +c}.
\label{eq10}
\end{equation}
Here $c$ is an integration constant. It follows from (\ref{eq04}) that $\epsilon_1=\eta=\rm const$ and $\varepsilon_2=0$, so this solution is trivial. 

In addition, one finds four nontrivial solutions to (\ref{eq00})\cite{Anguelova:2017djf}
\begin{eqnarray}
H_{1}(t)&=&-\frac{\beta}{\eta}\tan(\beta t+\gamma),\label{H1}\\
H_{2}(t)&=&\frac{\beta}{\eta}\cot(\beta t+\gamma),\label{H2}\\
H_{3}(t)&=&\frac{\beta}{\eta}\tanh(\beta t+\gamma),\label{H3}\\
H_{4}(t)&=&\frac{\beta}{\eta}\coth(\beta t+\gamma),\label{H4}
\end{eqnarray}
where $\beta$ and $\gamma$ are integration constants. 

\begin{figure}[h]
\begin{center}
\includegraphics[scale=0.6]{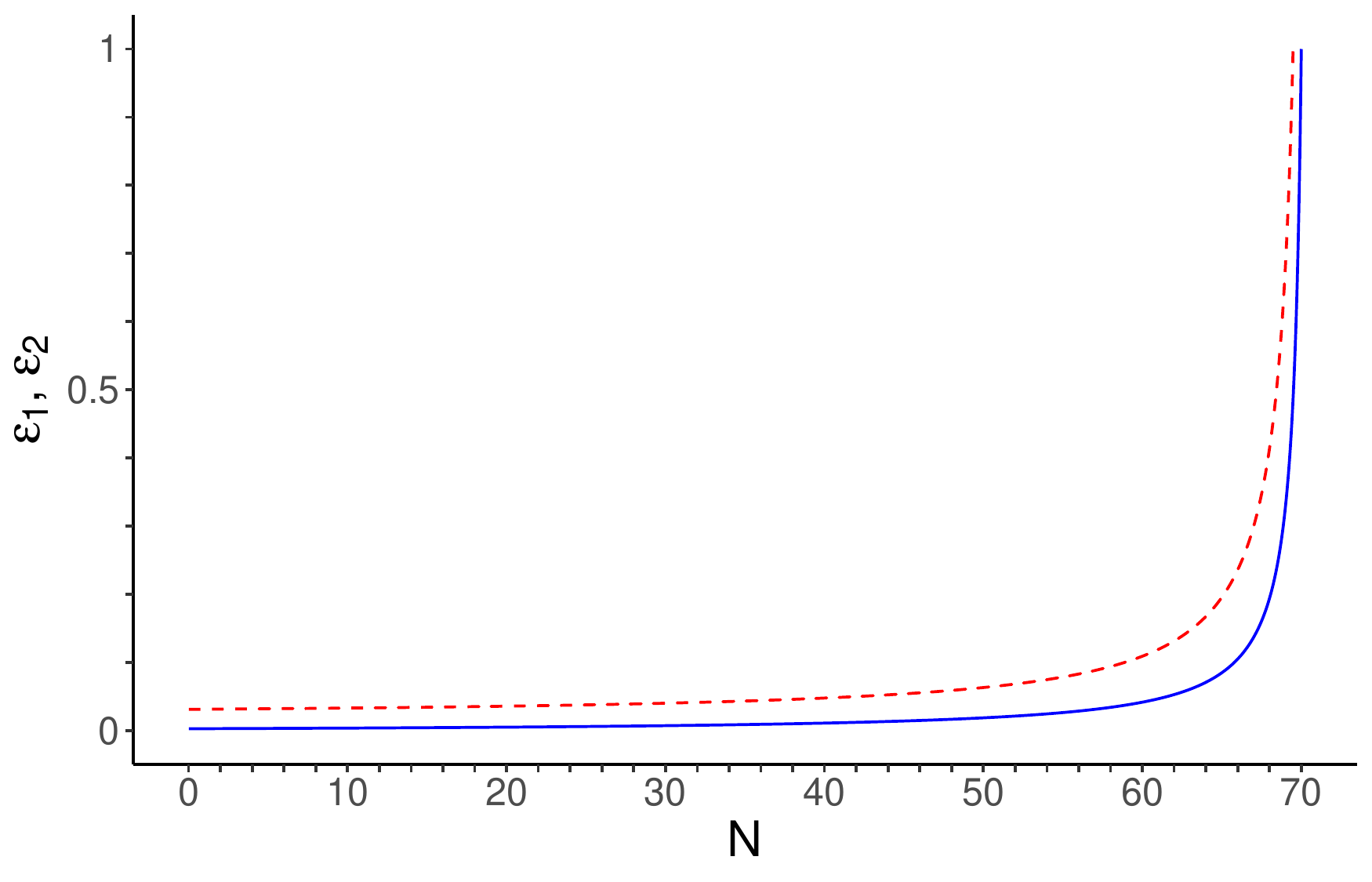}
\caption{The slow-roll parameters $\varepsilon_{1}$ (solid blue line) and $\varepsilon_{2}$ (dashed red line) as a function of the e-fold number $N$, for $\eta=-0.013$ and $N_{f}=70$.}
\label{plotEtaN}
\end{center}
\end{figure}

Using $H_{1}$ solution, a straightforward integration of (\ref{Ndef}) gives
\begin{equation}
N(t)=\frac{1}{\eta}\log\cos(\beta t+\gamma)+C,
\label{N}
\end{equation}
where $C$ is another integration constant. Using (\ref{eq1a}) and (\ref{eq1b}), from (\ref{H1}) we find
\begin{equation}
\varepsilon_{1}(t)=\frac{\eta}{\sin^{2}(\beta t+\gamma)},
\label{eq405}
\end{equation}
\begin{equation}
\varepsilon_{2}(t)=2\eta\cot^{2}(\beta t+\gamma).
\label{eq406}
\end{equation}
Then, combining (\ref{N}) with (\ref{eq405}) and (\ref{eq406}) we obtain
\begin{equation}
\varepsilon_{1}(N)=\frac{\eta}{1-e^{2\eta (N-C)}},
\label{e1N}
\end{equation}
\begin{equation}
\varepsilon_{2}(N)=\frac{2\eta e^{2\eta (N-C)}}{1-e^{2\eta (N-C)}}.
\label{e2N}
\end{equation}
From expressions (\ref{e1N}) and (\ref{e2N}), it is clear that the parameter $\eta$ must be positive. Inflation ends approximately at the end of the slow-role regime, i.e., when the first few horizon-flow parameters  $\varepsilon_i$ are close to 1. It is natural to assume that inflation ends at the point $t_{\rm f}$  where $\varepsilon_1(t_{\rm f})=1$ and $N=N_{\rm f}$. To fix the initial values $\varepsilon_{1\rm i}$ and $\varepsilon_{2\rm i}$, we assume $N=0$ at the begining of inflation ($t_{\rm i}=0$). From this condition, the constant $C$ can be calculated using (\ref{e1N})
\begin{equation}
C=N_{\rm f}-\frac{1}{2\eta}(1-\eta),
\end{equation}
and expressions for $\varepsilon_{1}(N)$ and $\varepsilon_{2}(N)$ become
\begin{equation}
\varepsilon_{1}(N)=\frac{\eta}{1-(1-\eta)e^{2\eta(N-N_{\rm f})}},
\label{e1Nf}
\end{equation}
\begin{equation}
\varepsilon_{2}(N)=\frac{2\eta(1-\eta)e^{2\eta(N-N_{\rm f})}}{1-(1-\eta)e^{2\eta(N-N_{\rm f})}}.
\label{e2Nf}
\end{equation}
In Fig. \ref{plotEtaN}, we plot\footnote{In order to display the results graphically, it is necessary to fix the value of the parameter $\eta$, which should be small (and negative, see Section 4), e.g., $\eta=-0.013$.} the evolution of the slow-roll parameters.

Note that $H_{2}$ solution is obtained from $H_{1}$ by replacing $\gamma\rightarrow\gamma+\pi$. Then, we find  
\begin{equation}
\varepsilon_{1}(t)=\frac{\eta}{\cos^{2}(\beta t+\gamma)},
\end{equation}
\begin{equation}
\varepsilon_{2}(t)=2\eta\tan^{2}(\beta t+\gamma),
\end{equation}
\begin{equation}
N(t)=\frac{1}{\eta}\log\sin(\beta t+\gamma)+C,
\end{equation}
which leads to the same expressions for parameters $\varepsilon_{1}(N)$ and $\varepsilon_{2}(N)$ as in the previous case, for $H_{1}$. 

Applying the same procedure for $H_{3}$  we obtain 
\begin{equation}
\varepsilon_{1}(t)=-\frac{\eta}{\sinh^{2}(\beta t+\gamma)},
\end{equation}
\begin{equation}
\varepsilon_{2}(t)=-2\eta\coth^{2}(\beta t+\gamma),
\end{equation}
\begin{equation}
N(t)=\frac{1}{\eta}\log\cosh(\beta t+\gamma)+C.
\end{equation}
In this case, we find the same expressions for $\varepsilon_{1}(N)$ and $\varepsilon_{2}(N)$ as in equations (\ref{e1Nf}) and (\ref{e2Nf}). 

The solution $H_{4}$ can not provide a valid model for inflation because the parameters 
\begin{equation}
\varepsilon_{1}(t)=\frac{\eta}{\cosh^{2}(\beta t+\gamma)},
\end{equation}
\begin{equation}
\varepsilon_{2}(t)=-2\eta\tanh^{2}(\beta t+\gamma),
\end{equation}
cannot be simultaneously positive. 

\begin{figure}[h!]
\begin{center}
\includegraphics[scale=0.75]{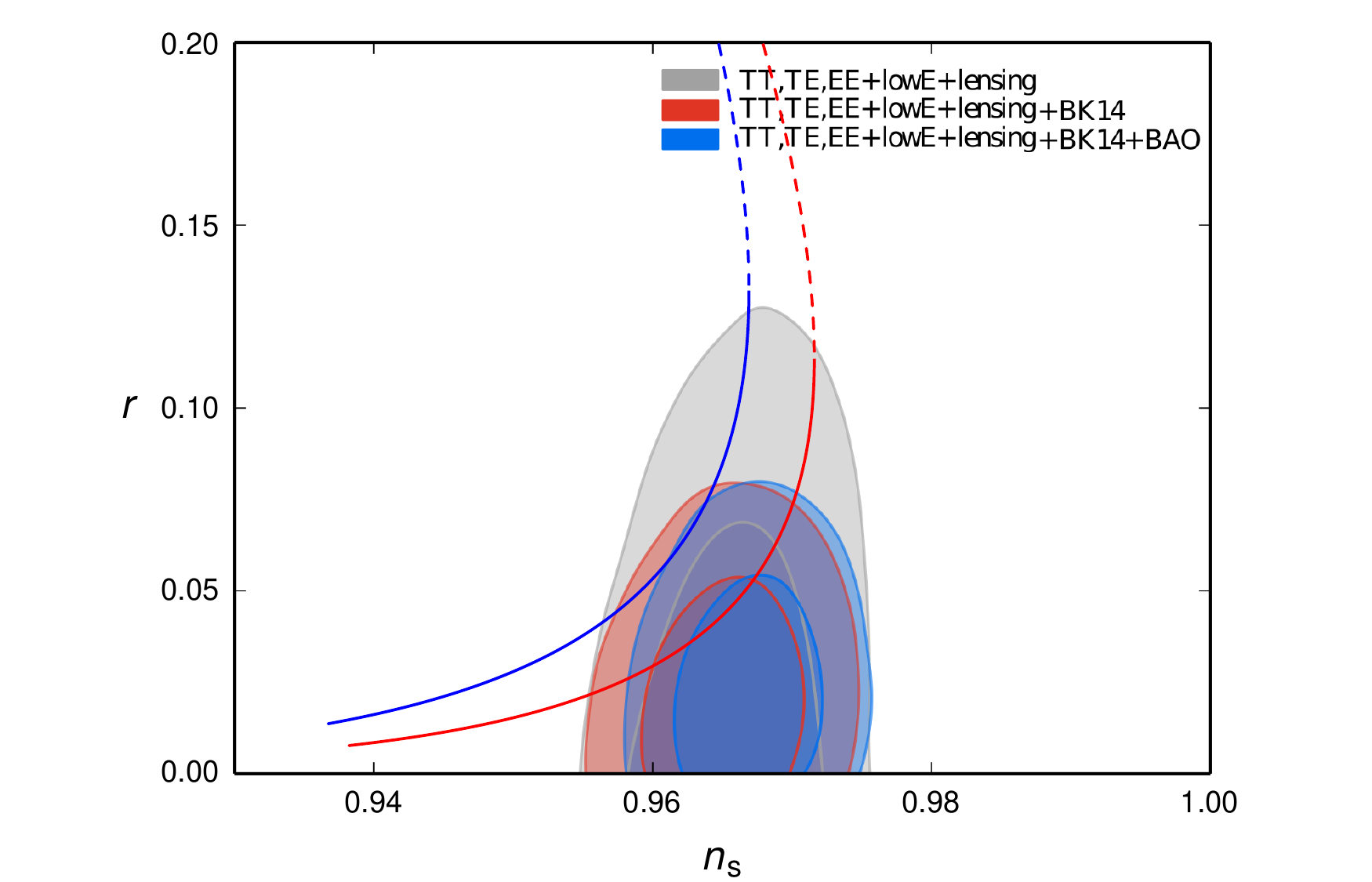}
\caption{$r$ versus $n_{\rm s}$ diagram with observational constraints from Planck mission\cite{Planck:2018jri}. Dashed line - prediction for models with $\eta>0$ (the solutions $H_{1}$ and $H_{2}$), $0<\eta< 0.01$, solid line - prediction for model  with $\eta<0$  (the solution $H_{3}$), $-0.03<\eta< 0$, for $N_{\rm f}=60$ (blue line) and $N_{\rm f}=70$ (red line).}
\label{nsr}
\end{center}
\end{figure}

\begin{figure}[h!]
\begin{center}
\includegraphics[scale=0.4]{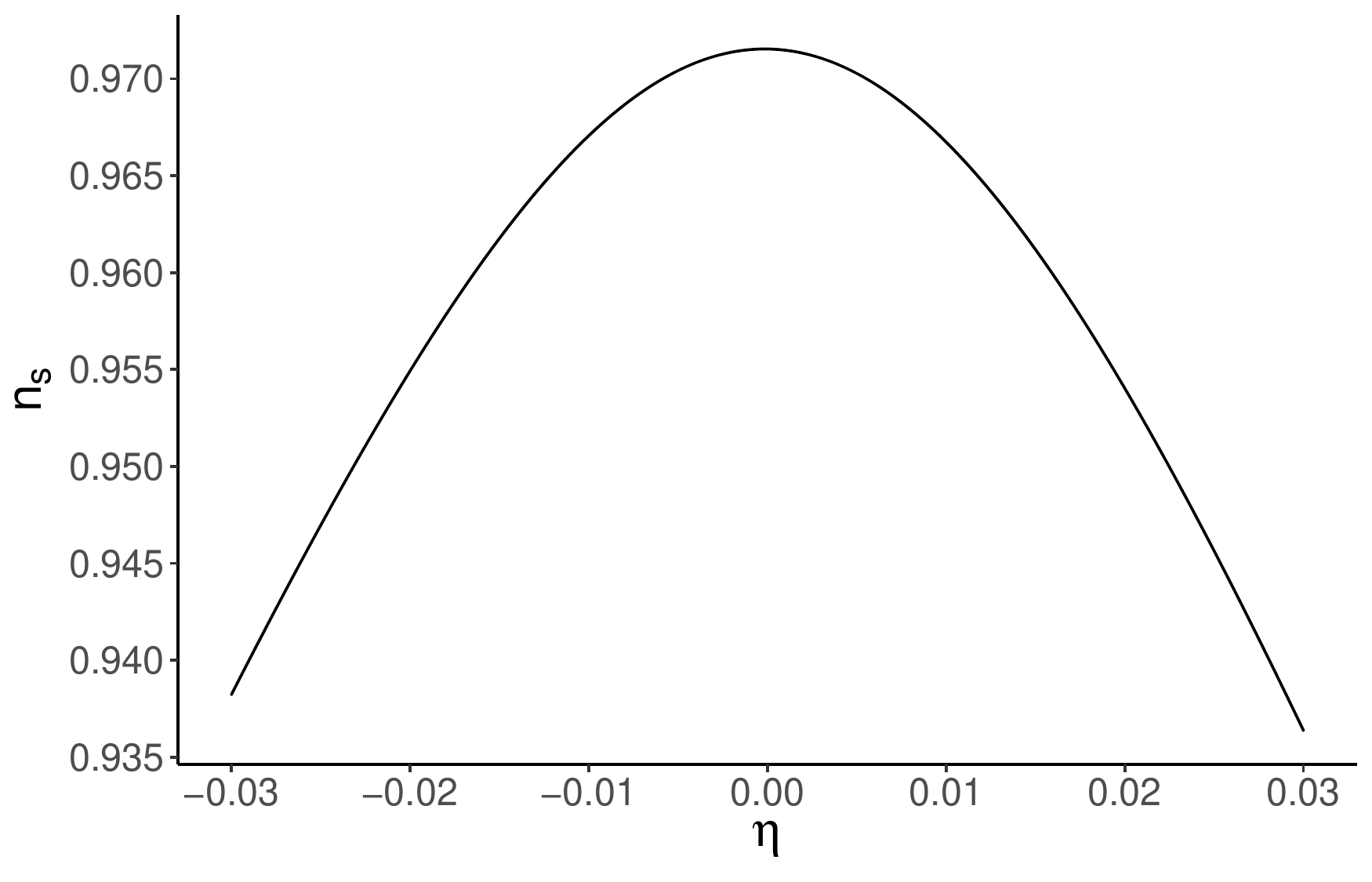}
\hspace{0.5cm}
\includegraphics[scale=0.4]{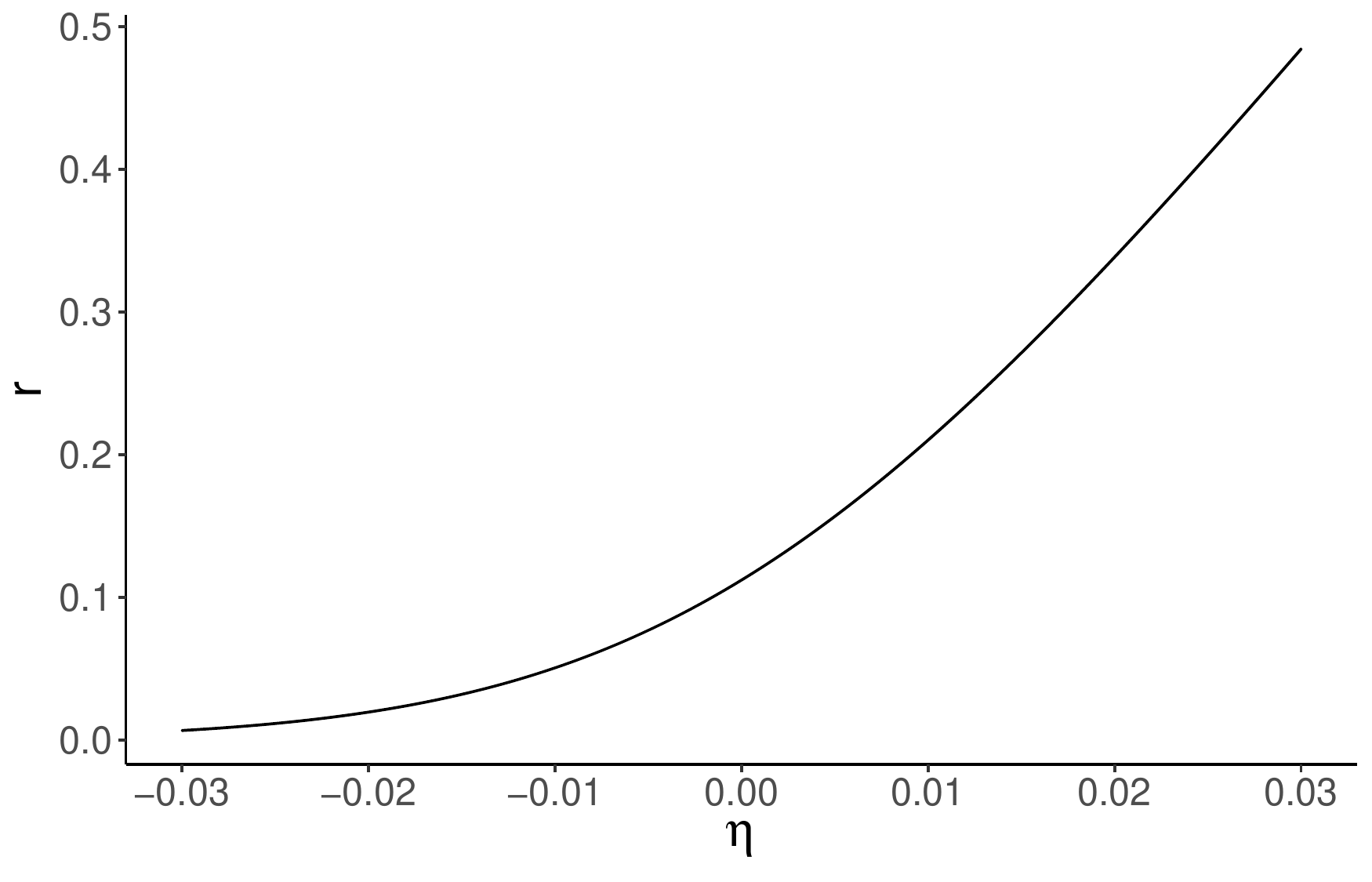}
\caption{$n_{\rm s}$ (left panel) and $r$ (right panel) versus $\eta$.}
\label{nsreta}
\end{center}
\end{figure}

In order to compare the prediction of our inflationary model with observational data, we need to calculate the observational parameters, such as the scalar spectral index $n_{\rm s}$ and the tensor-to-scalar ratio $r$. At the lowest order in the slow-roll parameters, these parameters are given by \cite{Steer:2003yu}
\begin{equation}
n_{\rm s}\simeq 1-2\varepsilon_{1\rm i}-\varepsilon_{2\rm i},
\end{equation}
\begin{equation}
r\simeq 16\varepsilon_{1\rm i},
\end{equation}
where $\varepsilon_{1\rm i}=\varepsilon_{1}(t_{\rm i})$ and $\varepsilon_{2\rm i}=\varepsilon_{2}(t_{\rm i}) $.
In Fig. \ref{nsr}, we depict the results in $n_{\rm s}$-$r$  plane. 
Clearly, a better agreement with observations is obtained with the solutions in which the values of $\eta$ are negative. The dependence of the parameters $n_{\rm s}$ and $r$ on $\eta$ is shown in Fig. \ref{nsreta}, from which one can conclude that the agreement with observations is better for negative and small $\eta$.

As the solutions in this section are model-independent, they could provide a feasible inflationary scenario in any model that satisfies the condition (\ref{defeta}) with constant $\eta$. We demonstrate this in the following section.

\section{Constant-roll inflation with tachyon matter}

Here we analyze the constant-roll inflation with constant $\eta$, with dynamics described by the tachyon field, in the framework of the RSII cosmology and standard cosmology. Our model is based on a braneworld scenario in which our universe can be considered as a brane, i.e., a four-dimensional hypersurface embedded in a higher-dimensional spacetime. The RSII model \cite{Randall:1999vf, Dimitrijevic:2018nlc} describes a universe containing two branes with opposite tensions, separated in the fifth dimension, where only gravity can propagate. Observers reside on the positive tension brane, and the negative tension brane is pushed off to infinity.  

\subsection{The standard and the RSII cosmology with tachyon matter }

In a flat FLRW universe in standard cosmology, the Friedmann equations are of the form\cite{Steer:2003yu}

\begin{equation}
H^{2}=\frac{8\pi}{3M_{4}^2}\rho,
\label{eqfs1}
\end{equation}
\begin{equation}
\dot{H}=-\frac{4\pi}{M_{4}^2}(\rho+p).
\label{eqfs2}
\end{equation}
On the other hand, the Universe expansion in RSII cosmology\cite{Randall:1999vf} is described by modified Friedmann equations\cite{Mohammadi:2020ftb, Milosevic:2018gck}
\begin{equation}
H^{2}=\frac{8\pi}{3M_{4}^2}\rho(1+\frac{\rho}{2\lambda}),
\label{F1}
\end{equation}
\begin{equation}
\dot{H}=-\frac{4\pi}{M_{4}^2}(1+\frac{\rho}{\lambda})(\rho+p),
\label{F2}
\end{equation}
which apparently differ from the standard Friedmann equations. The parameter $\lambda$ denotes the brane tension, which is related to the five-dimensional and four-dimensional Planck masses, $M_{5}$ and $M_{4}$, respectively:
\begin{equation}
\lambda=\frac{3}{4\pi}\left(\frac{M_{5}^3}{M_{4}}\right)^2.  
\end{equation}
Following Ref. \citen{Mohammadi:2020ftb}, we consider the case when the energy density is much larger than the tension of the brane, i.e., $\rho\gg\lambda$. As a consequence, equations (\ref{F1}) i (\ref{F2}) are simplified:
\begin{equation}
H^{2}\simeq\frac{4\pi}{3M_{4}^2}\frac{\rho^2}{\lambda},
\label{F1approx}
\end{equation}
\begin{equation}
\dot{H}\simeq -\frac{4\pi}{M_{4}^2}\frac{\rho}{\lambda}(\rho+p).
\label{F2approx}
\end{equation}

The dynamics of a tachyon field $\theta$ is described by a Lagrangian of the Dirac-Born-Infeld (DBI) form\cite{Sen:1999md}. In a homogeneous and isotropic background, the Lagrangian can be put in the form
\begin{equation}
{\cal L}=-V(\theta)\sqrt{1-\dot{\theta}^2}.
\end{equation}
The energy density and pressure of the tachyon field are given by\cite{Steer:2003yu} 
\begin{equation}
\rho=\frac{V}{\sqrt{1-\dot{\theta}^2}}, \label{eqrho}  
\end{equation}
\begin{equation}
p=-V\sqrt{1-\dot{\theta}^2}.
\label{eqp} 
\end{equation}
Using the Hamilton-Jacobi formalism, we express the Hubble expansion rate as a function of the tachyon field $H=H(\theta)$, and the time derivative of $H$ via $\dot{H}=H_{,\theta}\dot{\theta}$, where $H_{,\theta}$ denotes a derivative of $H$ respect to $\theta$. Then, combining Friedmann’s equations, (\ref{eqfs1}) and (\ref{eqfs2}), or (\ref{F1approx}) and (\ref{F2approx}), with (\ref{eqrho}) and (\ref{eqp}) one obtains   
\begin{equation}
\dot{\theta}=-\frac{n}{3}\frac{H_{,\theta}}{H^{2}}.
\label{dottheta} 
\end{equation}
Here, the integer $n$ can take two values: $n=1$ for RSII cosmology and $n=2$ for standard cosmology. As expected, the expression for $\dot{\theta}$ differs from the expression in a model with a canonical scalar field\cite{Mohammadi:2020ftb} $\phi$
\begin{equation}
\dot{\phi}=-\frac{M_{5}^{3}}{4\pi}\frac{H_{,\phi}}{H},
\end{equation}
suggesting that the model with a tachyon field may give a different prediction for the observational parameters.

\subsection{The constant-roll inflation with a tachyon field }

Next, we calculate the observational parameters $n_{\rm s}$ and $r$. Equation (\ref {eq00}), using (\ref{dottheta}), can be transformed to a differential equation with respect to the tachyon field $\theta$
\begin{equation}
H_{,\theta\theta}H-H_{,\theta}^2-3\frac{\eta}{n} H^{4}=0,
\label{difH}
\end{equation}
with the solution of the form
\begin{equation}
H(\theta)=\frac{2n C_{1}e^{\sqrt{C_{1}}(\theta+C_{2})}}{e^{2\sqrt{C_{1}}(\theta+C_{2})}-3\bar{\eta} C_{1}},
\label{sol}
\end{equation}
where $\bar{\eta}=n\eta$. The integration constants  $C_{1}$ i $C_{2}$, in the expression (\ref{sol}), can be absorbed by rescaling $H$, $\theta$ and $\bar{\eta}$. It is easy to check that, without loss of generality, we can set $C_{1}=1$ and $C_{2}=0$, yielding
\begin{equation}
H=\frac{2n e^{\theta}}{e^{2\theta}-3\bar{\eta}}.
\label{Htheta}
\end{equation}

One can combine (\ref{dottheta}) and (\ref{Htheta}) to find the time dependence of $\theta$ and $H$. First, one finds a simple expression
\begin{equation}
\dot{\theta}=\frac16 (e^\theta+3\bar{\eta} e^{-\theta}).
\end{equation}
This may easily be integrated, yielding
\begin{eqnarray}
e^\theta = \sqrt{3\bar{\eta}}\, \tan \left( \sqrt{\bar{\eta}/12}\, t +C_3\right),\quad \bar{\eta}>0,\label{etheta}\\
e^\theta = -\sqrt{3|\bar{\eta}|}\, \tanh \left( \sqrt{|\bar{\eta}|/12}\, t +C_3\right),\quad \bar{\eta}<0\label{etheta1}.
\end{eqnarray}
Plugging (\ref{etheta}) and (\ref{etheta1}) in (\ref{Htheta}) one finds
\begin{eqnarray}
H(t)&=& -\frac{n}{\sqrt{3\bar{\eta}}} \tan ( \sqrt{\bar{\eta}/3}\, t +2C_3),\quad \bar{\eta}>0,
\label{eq12}\\
H(t)&=& -\frac{n}{\sqrt{3|\bar{\eta}|}} \tanh ( \sqrt{|\bar{\eta}|/3}\, t +2C_3),\quad \bar{\eta}<0.
\label{eq1212}
\end{eqnarray}
Using these expressions for $H$, the slow-roll parameters become
\begin{eqnarray}
\varepsilon_1(t)&=& \frac{\bar{\eta}}{n\sin^2 \left( \sqrt{\bar{\eta}/3}\, t +2C_3\right)} ,\quad \bar{\eta}>0,
\label{epslion1N}\\
\varepsilon_1(t) &=& \frac{ |\bar{\eta}|}{n\sinh^2\left( \sqrt{|\bar{\eta}|/3}\, t +2C_3\right)},\quad \bar{\eta}<0.
\label{epslion2N}
\end{eqnarray}
Integration constant $C_{3}$ is fixed by the initial value of $\varepsilon_{1\rm i}$. According to (\ref{e1Nf}) $\varepsilon_{1\rm i}$ depends on values of $\bar{\eta}$ and $N_{\rm f}$. Obviously, equations (\ref{eq12}) and (\ref{eq1212}) agree with  the more general solutions,  (\ref{H1}) and (\ref{H3}) respectively, with $\beta=\sqrt{|\bar{\eta}|/3}$ (and $\gamma=2C_{3}$). This confirms that the results (\ref{eq12}) and (\ref{epslion2N}) are independent of the chosen values of constants $C_{1}$ and $C_{2}$.

Integrating $H(t)$, we obtain the time evolution of the scale factor 
\begin{eqnarray}
a(t)&\propto&\left[\cos\left(\sqrt{\bar{\eta}/3}\,t+2C_3\right)\right]^{\frac{n}{\bar{\eta}}},\quad \eta>0,\\
a(t)&\propto&\left[\cosh\left(\sqrt{|\bar{\eta}|/3}\,t+2C_3\right)\right]^{-\frac{n}{|\bar{\eta}|}},\quad \eta<0,
\end{eqnarray}
which is consistent with the results of Ref. \citen{Anguelova:2017djf}. 

\begin{figure}[h]
\begin{center}
\includegraphics[scale=0.4]{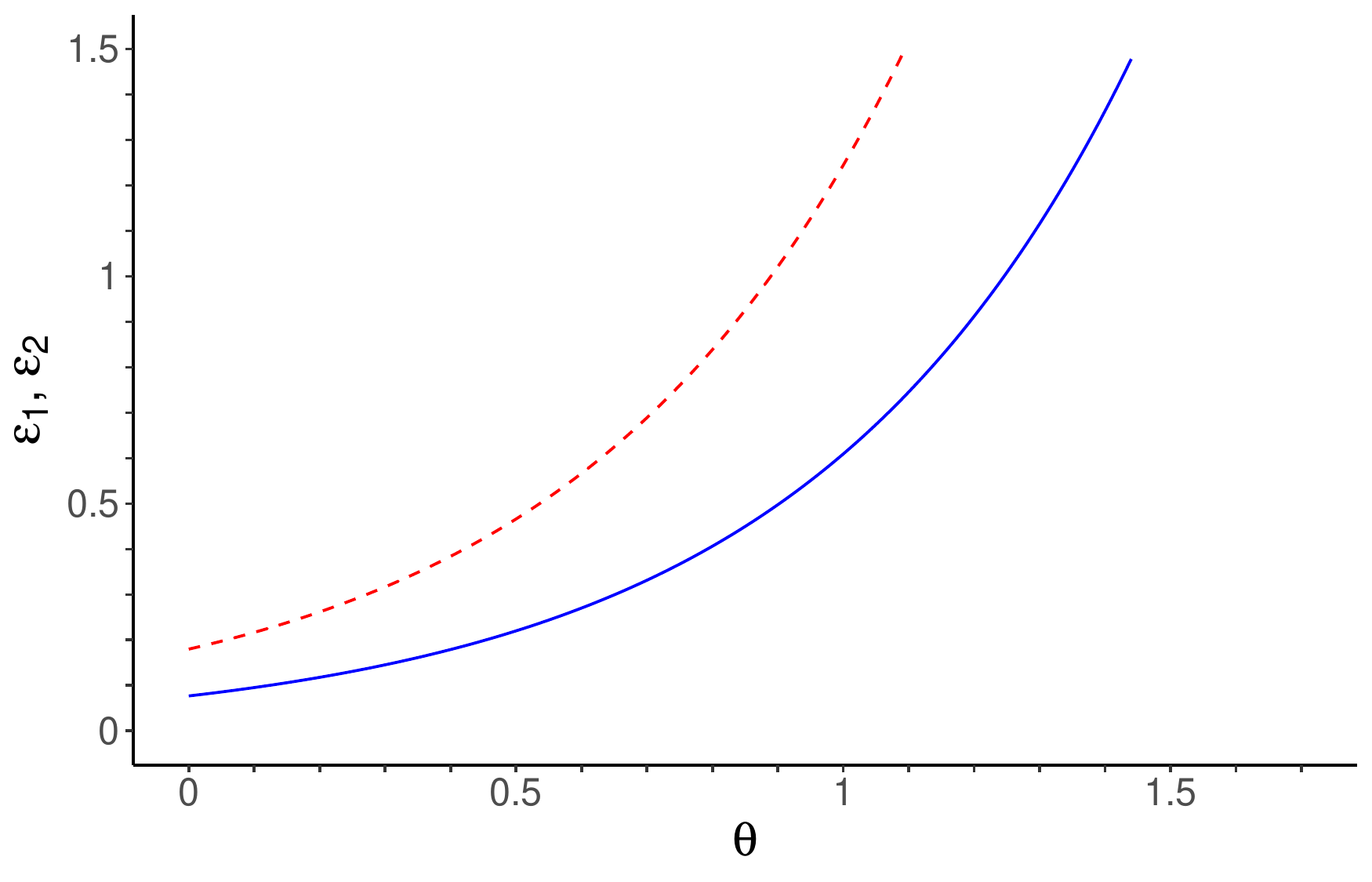}
\caption{The slow-roll parameters $\varepsilon_{1}$  (solid blue line) and $\varepsilon_{2}$ (dashed red line) versus $\theta$ during inflation in RSII cosmology for $\bar{\eta}=-0.013$.}
\label{figepsilon}
\end{center}
\end{figure}

Plugging (\ref{Htheta}) in (\ref{eq1a}) and (\ref{eq1b}) and  utilizing (\ref{dottheta}) we obtain the expressions for the slow-roll parameters 
\begin{eqnarray}
\varepsilon_{1}(\theta)&=&-\frac{\dot{H}}{H^2}=\frac{1}{3}\left(\frac{H_{,\theta}}{H^2}\right)^{2}=\frac{1}{12}e^{-2\theta}\left(e^{2\theta}+3\bar{\eta}\right)^2,\label{eps1}\\
\varepsilon_{2}(\theta)&=&\frac{1}{6}e^{-2\theta}\left(e^{2\theta}-3\bar{\eta}\right)^2\label{eps2}.
\end{eqnarray}
In Fig. \ref{figepsilon} we plot $\varepsilon_{1}$ and $\varepsilon_{2}$ as functions of $\theta$.

The parameter $\bar{\eta}$ can be positive or negative. It is obvious from equation (\ref{Htheta}) that $H>0$ when $\bar{\eta}<0$. To prove that $H>0$ when $\bar{\eta}>0$ too, note that $\varepsilon_{1}>\bar{\eta}$. Solving (\ref{eps1}) as a quadratic equation for $e^{\theta}$ we obtain two solutions, $f_{1}$ and $f_{2}$, as functions of $\bar{\eta}$. The solution $f_{1}$ ($f_{1}>f_{2}$) satisfies the inequality $e^{2\theta}>3\bar{\eta}$, so $H>0$. The other solution ($f_{2}$) leads to an unphysical  Hubble expansion rate.

\begin{figure}[h]
\begin{center}
\includegraphics[scale=0.4]{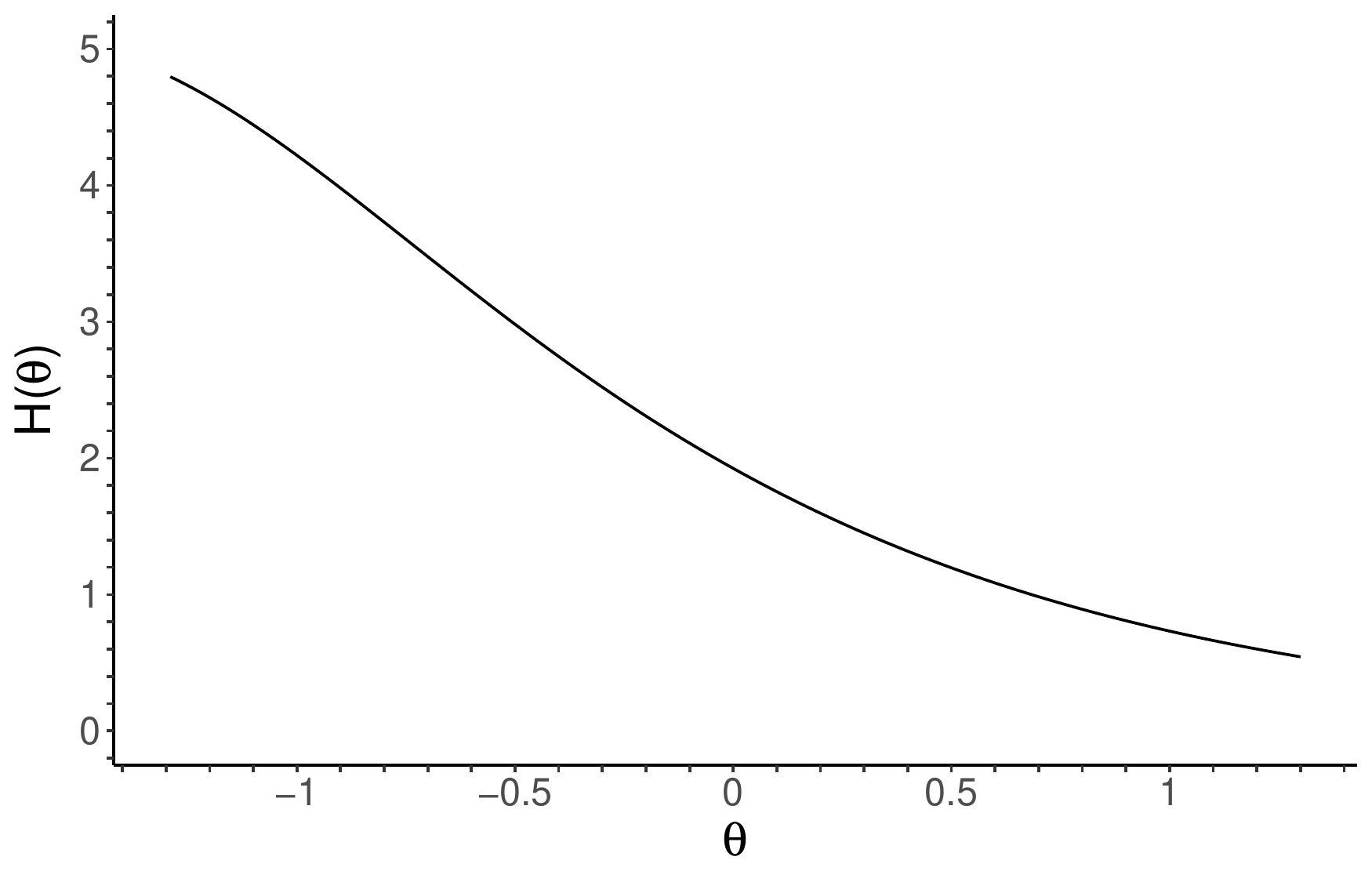}
\includegraphics[scale=0.4]{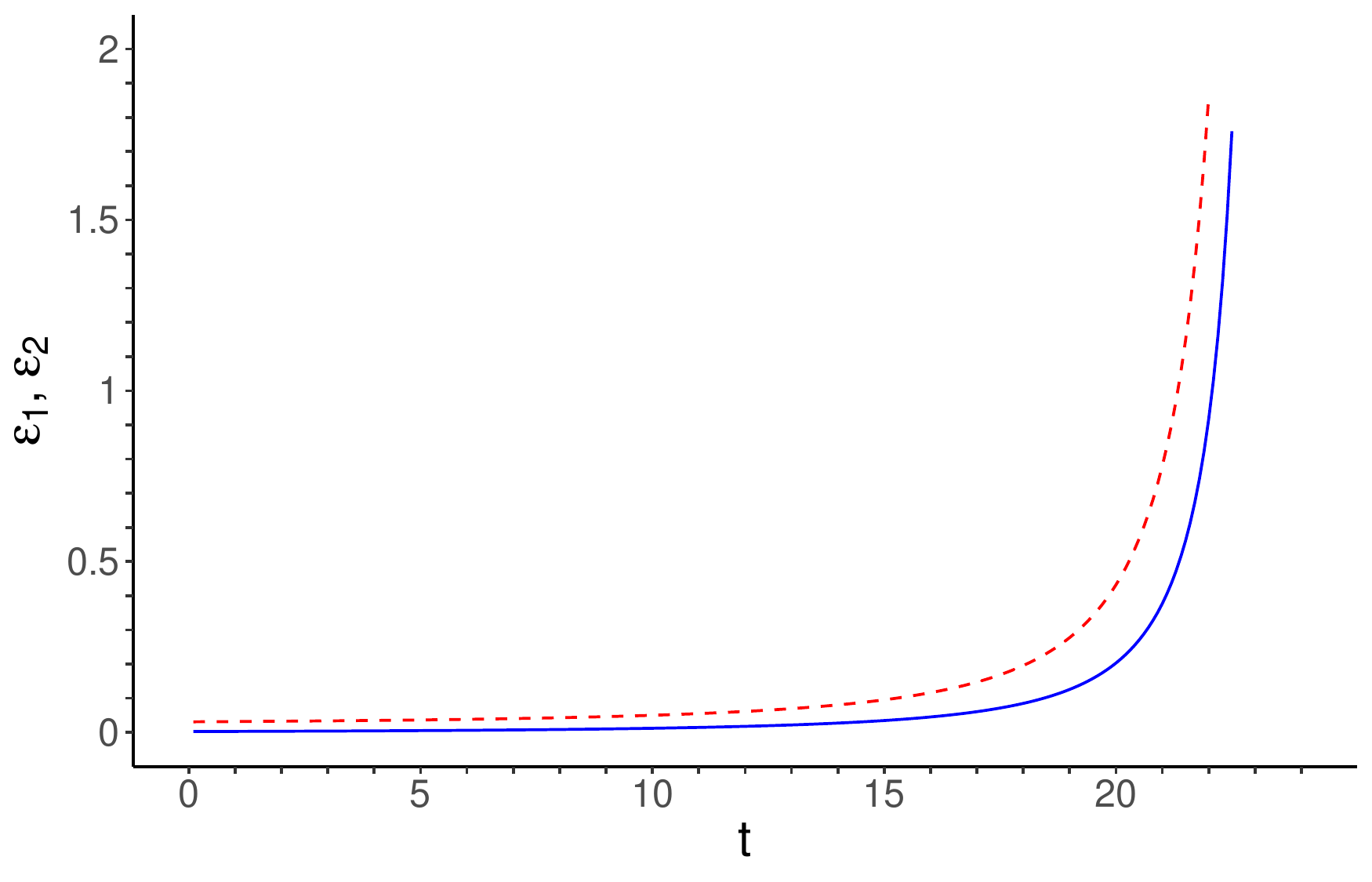}
\caption{The Hubble parameter versus the tachyon field (left panel) and the behavior of the slow-roll parameters $\varepsilon_{1}$ (solid blue line) and $\varepsilon_{2}$ (dashed red line) during the inflationary times versus tachyon field (right panel), for $\bar{\eta} = -0.013$.}
\label{plotHplotE}
\end{center}
\end{figure}

Let us consider our model in RSII cosmology for $H>0$ for $N_{f}=70$ and $\bar{\eta}=-0.013$. We prove that $H>0$ for any value of $\theta$. From (\ref{e1Nf}) we obtain $\varepsilon_{1\rm i}=0.0025$. Then, solving (\ref{eps1}) as a quadratic equation for $\theta$ we find the initial value $\theta_{\rm i}=-1.2$. Using (\ref{Htheta}), we find the corresponding value of the Hubble expansion rate $H_{\rm i}=4.64$. Apparently, the field is negative at the beginning of inflation while the Hubble expansion rate is always greater than zero (see Fig. \ref{plotHplotE}).
The end-value of the field $\theta_{\rm f}$ is obtained from the condition for the end of inflation $\varepsilon_{1}(t_{\rm f})=1$. By making use of (\ref{eps1}) we find 
\begin{equation}
\theta_{\rm f}=\ln\left(\sqrt{3}\sqrt{2+2\sqrt{1-\eta}-\eta}\right). 
\label{thetaf}
\end{equation}
For the model with $\bar{\eta}=-0.013$, inflation ends at $t_{\rm f}=22.07$ (see Fig. \ref{plotHplotE}) with $H_{\rm f}=0.57$, when the field reaches the value $\theta_{\rm f}=1.24$. 

The observational parameters for the inflation model, driven by the tachyon field, have already been calculated in standard cosmology\cite{Steer:2003yu} and RSII cosmology\cite{Bilic:2016fgp}. The expressions for the scalar spectral index and the tensor-to-scalar ratio, at second order in the slow-roll parameters, read
\begin{equation}
n_{\rm s}=1-2\varepsilon_{1\rm i}-\varepsilon_{2\rm i}-\left(2\varepsilon_{1\rm i}^2+(2C'+3-2\alpha)\varepsilon_{1\rm i}\varepsilon_{2\rm i}+C'\varepsilon_{2\rm i}\varepsilon_{3\rm i}\right),
\end{equation}
\begin{equation}
r=16\varepsilon_{1\rm i}\left(1+C'\varepsilon_{2\rm i}-2\alpha\varepsilon_{1\rm i}\right).
\end{equation}
The value of the parameter $\alpha$ differs in standard ($\alpha=1/6$) and RSII cosmology ($\alpha=1/12$), while $C'=-0.72$. The distinction of the constant-roll inflation is also reflected in the value of the third slow-roll parameter $\varepsilon_{3}$. According to (\ref{parrec1}), $\varepsilon_{3\rm i}=2\varepsilon_{1\rm i}$, that holds only for the constant-roll inflation, i.e., for the model with $\eta$ constant. 

\begin{figure}[h]
\begin{center}
\includegraphics[scale=0.75]{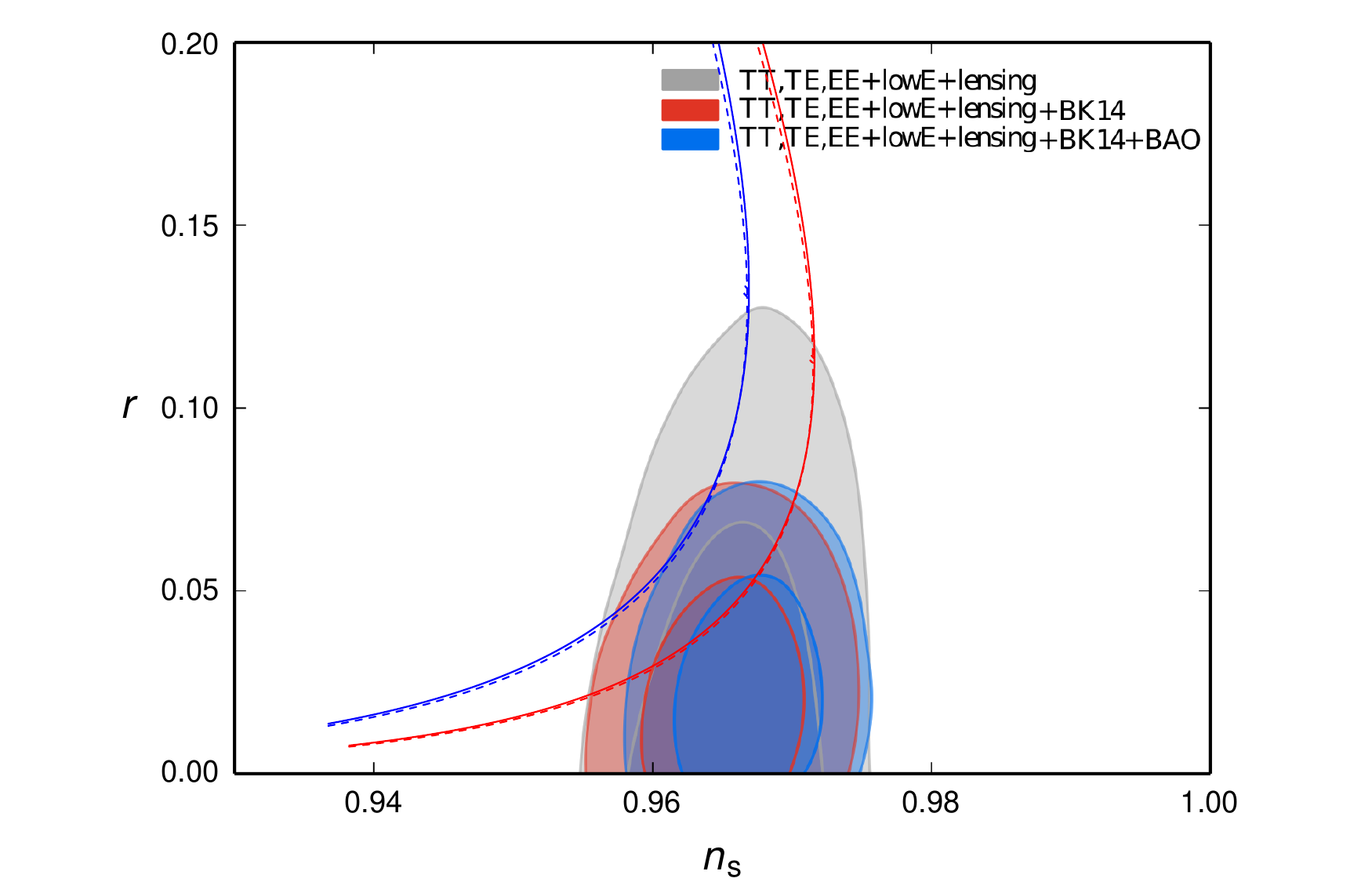}
\caption{$r$ versus $n_{\rm s}$ diagram with observational constraints from Planck mission\cite{Planck:2018jri}. The analytical results are depicted for fixed $N_{\rm f}=60$ (blue lines) and $N_{\rm f}=70$ (red lines). The solid and dashed lines are obtained from the expressions for $n_{\rm s}$ and $r$ up to the first and the second order in the slow-role parameters in RSII cosmology, respectively. The parameter $\eta$ varies along the lines in the interval $-0.03<\eta<0.03$.}
\label{ns_r2}
\end{center}
\end{figure}

As in the standard cosmology, the RSII inflationary model with a tachyon field is fully analytical. In Fig. \ref{ns_r2}, we present the results for $n_{\rm s}$ and $r$ parameters, superimposed on the observational constraints. A better agreement of analytical and observational results is evident for a bit higher values of $N$, e.g., $N_{\rm f}=70$. It may be noted that the influence of the second order in the slow-roll parameters is insignificant. Finally, note that the difference between results in the standard tachyon inflationary model and the RSII inflationary model in the constant-roll inflation approach is small. 

\begin{figure}[h]
\begin{center}
\includegraphics[scale=0.5]{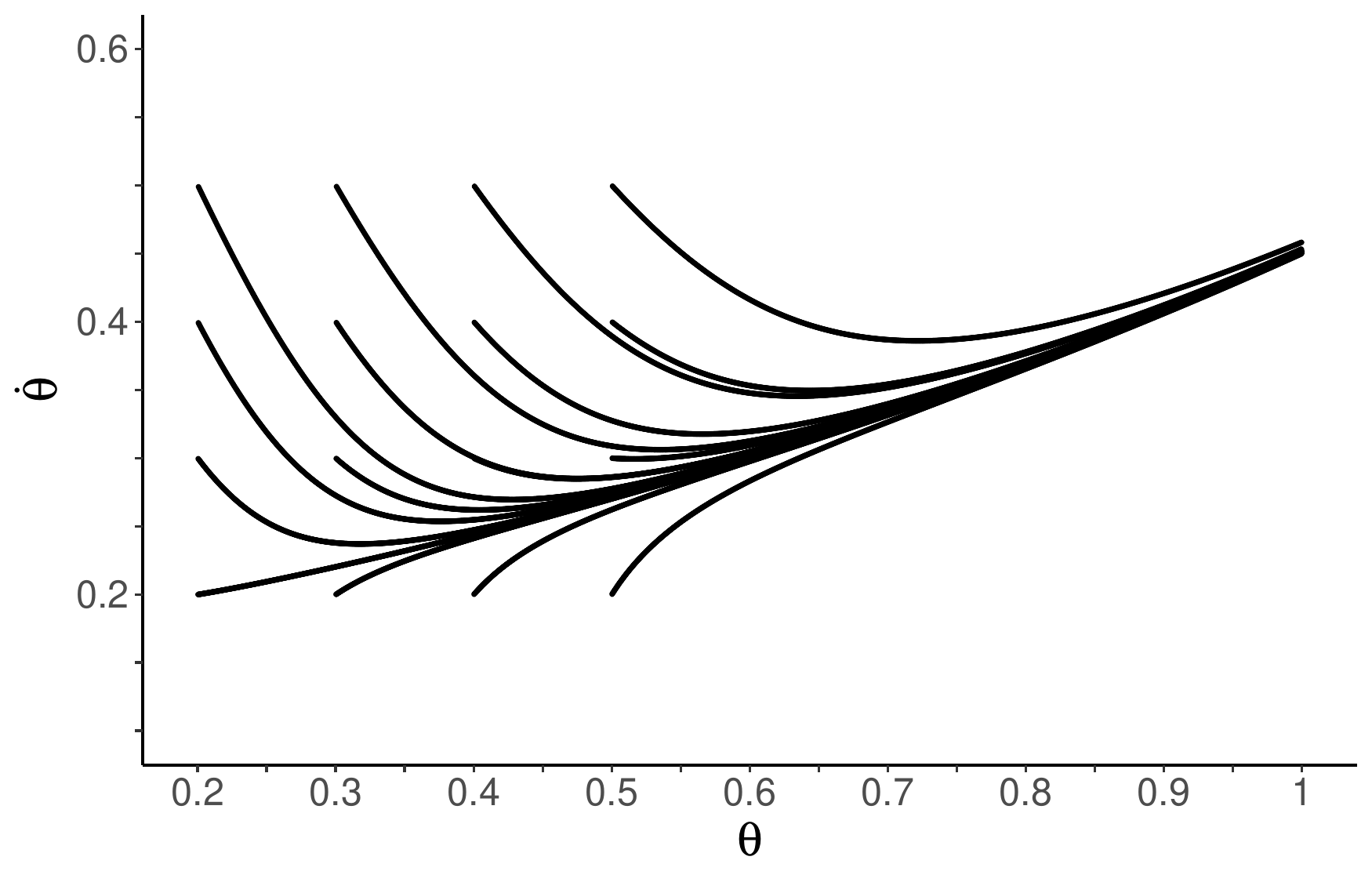}
\caption{The phase space trajectories (right panel) obtained for some initial values in range $0.2\leq\theta_{\rm i}\leq0.5$ and $0.2\leq\dot{\theta}_{\rm i}\leq0.5$, for the model with $\bar{\eta}=-0.013$ in units where $3M_{5}^{3}/(4\pi)=1$.}
\label{phase space}
\end{center}
\end{figure}

The attractor behavior of the solution is a necessary condition for a successful inflation model. Our model possesses a good attractor behavior, as presented in Fig. \ref{phase space}. Details will be presented elsewhere. 

\section{Summary and conclusions}

We have studied the constant-roll inflation in the framework of the RSII cosmology with a tachyon field. The slow-roll parameter $\eta$  with a fixed constant value leads to a differential equation for the Hubble expansion rate. This equation has exact analytical solutions. We have calculated the Hubble slow-roll parameters $\varepsilon_{i}$ as a function of $\eta$ for all four nontrivial solutions $H(\theta)$. It has been shown that three of four solutions for $H(\theta)$ provide a consistent inflationary model. Futhermore, all solutions lead to the same functions $\varepsilon_{1}(N)$ and $\varepsilon_{2}(N)$. 

We have calculated the values of the scalar spectral index $n_{\rm s}$ and the tensor-to-scalar ratio $r$. Besides, we have compared these values with the observation data. Using this comparison, we have estimated the values of parameter $\eta$. A better agreement is achieved for negative and small values of the $\eta$. In addition, we have calculated the observational parameters for standard and RSII cosmology at second order in the slow-roll parameters. No significant difference is obtained in these two cases. As a straightforward extension of this work, it would be of interest to apply the formalism of the constant-roll inflation to the holographic RSII model with a tachyon field \cite{Bilic:2018uqx}.

\section*{Acknowledgments}

This work has been supported by the ICTP-SEENET-MTP project NT-03 Cosmology-Classical and Quantum Challenges and the COST Action CA18108 "Quantum gravity phenomenology in the multi-messenger approach". M. Stojanovic acknowledges the support provided by the Serbian Ministry for Education, Science, and Technological Development under contract 451-03-47/2023-01/2000113. D. D. Dimitrijevic, G. S. Djordjevic, and M. Milosevic acknowledge the support provided by the Serbian Ministry for Education, Science, and Technological Development under contract 451-03-47/2023-01/2000124. In addition, G. S. Djordjevic acknowledges the support of the CEEPUS Program RS-1514-03-2223 "Gravitation and Cosmology" and the hospitality of the colleagues at the University of Banja Luka.



\begin{thebibliography}{00}  


\bibitem{Guth:1980zm}
A.~H.~Guth,
Phys. Rev. D \textbf{23} (1981), 347-356
doi:10.1103/PhysRevD.23.347

\bibitem{Feinstein:2002aj}
A.~Feinstein,
Phys. Rev. D \textbf{66} (2002), 063511
doi:10.1103/PhysRevD.66.063511
[arXiv:hep-th/0204140 [hep-th]].

\bibitem{Randall:1999vf}
L.~Randall and R.~Sundrum,
Phys. Rev. Lett. \textbf{83}, 4690-4693 (1999)
doi:10.1103/PhysRevLett.83.4690
[arXiv:hep-th/9906064 [hep-th]].

\bibitem{Mohammadi:2020ftb}
A.~Mohammadi, T.~Golanbari, S.~Nasri and K.~Saaidi,
Phys. Rev. D \textbf{101}, no.12, 123537 (2020)
doi:10.1103/PhysRevD.101.123537
[arXiv:2004.12137 [gr-qc]].

\bibitem{Bilic:2017orf}
N.~Bili\'c, S.~Domazet and G.~Djordjevic,
Class. Quant. Grav. \textbf{34} (2017) no.16, 165006
doi:10.1088/1361-6382/aa7e0f
[arXiv:1704.01072 [gr-qc]].

\bibitem{Bilic:2013dda}
N.~Bili\'c and G.~B.~Tupper,
doi:10.2478/s11534-013-0325-y
[arXiv:1309.6588 [hep-th]].

\bibitem{Yi:2017mxs}
Z.~Yi and Y.~Gong,
JCAP \textbf{03}, 052 (2018)
doi:10.1088/1475-7516/2018/03/052
[arXiv:1712.07478 [gr-qc]].

\bibitem{Mohammadi:2018oku}
A.~Mohammadi, K.~Saaidi and T.~Golanbari,
Phys. Rev. D \textbf{97}, no.8, 083006 (2018)
doi:10.1103/PhysRevD.97.083006
[arXiv:1801.03487 [hep-ph]].

\bibitem{Nguyen:2021emx}
D.~H.~Nguyen, T.~M.~Pham and T.~Q.~Do,
Eur. Phys. J. C \textbf{81} (2021) no.9, 839
doi:10.1140/epjc/s10052-021-09652-1
[arXiv:2107.14115 [gr-qc]].

\bibitem{Planck:2018jri}
Y.~Akrami \textit{et al.} [Planck],
Astron. Astrophys. \textbf{641}, A10 (2020)
doi:10.1051/0004-6361/201833887
[arXiv:1807.06211 [astro-ph.CO]].

\bibitem{Liddle:1994dx}
A.~R.~Liddle, P.~Parsons and J.~D.~Barrow,
Phys. Rev. D \textbf{50}, 7222-7232 (1994)
doi:10.1103/PhysRevD.50.7222
[arXiv:astro-ph/9408015 [astro-ph]].

\bibitem{Steer:2003yu}
D. A. Steer and F. Vernizzi,
Phys. Rev. D \textbf{70}, 043527 (2004)
doi:10.1103/PhysRevD.\\70.043527 
[arXiv:hep-th/0310139 [hep-th]].

\bibitem{Schwarz:2001vv}
D.~J.~Schwarz, C.~A.~Terrero-Escalante and A.~A.~Garcia,
Phys. Lett. B \textbf{517}, 243-249 (2001)
doi:10.1016/S0370-2693(01)01036-X
[arXiv:astro-ph/0106020 [astro-ph]].

\bibitem{Tsamis:2003px}
N.~C.~Tsamis and R.~P.~Woodard,
Phys. Rev. D \textbf{69}, 084005 (2004)
doi:10.1103/PhysRevD.69.084005
[arXiv:astro-ph/0307463 [astro-ph]].

\bibitem{Motohashi:2014ppa}
H.~Motohashi, A.~A.~Starobinsky and J.~Yokoyama,
JCAP \textbf{09}, 018 (2015)
doi:10.1088/1475-7516/2015/09/018
[arXiv:1411.5021 [astro-ph.CO]].

\bibitem{Martin:2012pe}
J.~Martin, H.~Motohashi and T.~Suyama,
Phys. Rev. D \textbf{87}, no.2, 023514 (2013)
doi:10.1103/PhysRevD.87.023514
[arXiv:1211.0083 [astro-ph.CO]].

\bibitem{Gao:2018tdb}
Q.~Gao, Y.~Gong and Q.~Fei,
JCAP \textbf{05}, 005 (2018)
doi:10.1088/1475-7516/2018/05/005
[arXiv:1801.09208 [gr-qc]].

\bibitem{Anguelova:2017djf}
L.~Anguelova, P.~Suranyi and L.~C.~R.~Wijewardhana,
JCAP \textbf{02}, 004 (2018)
doi:10.1088/1475-7516/2018/02/004
[arXiv:1710.06989 [hep-th]].

\bibitem{Dimitrijevic:2018nlc}
D.~D.~Dimitrijevic, N.~Bili\'c, G.~S.~Djordjevic, M.~Milosevic and M.~Stojanovic,
Int. J. Mod. Phys. A \textbf{33} (2018) no.34, 1845017
doi:10.1142/S0217751X18450173

\bibitem{Milosevic:2018gck}
M.~Milo\v{s}evi\'c, N.~Bili\'c, D.~D.~Dimitrijevi\'c, G.~S.~Djordjevi\'c and M.~Stojanovi\'c,
AIP Conf. Proc. \textbf{2075}, no.1, 090009 (2019)
doi:10.1063/1.5091223
[arXiv:1809.04939 [gr-qc]].

\bibitem{Sen:1999md}
A.~Sen,
JHEP \textbf{10}, 008 (1999)
doi:10.1088/1126-6708/1999/10/008
[arXiv:hep-th/9909062 [hep-th]].

\bibitem{Bilic:2016fgp}
N.~Bilic, D.~Dimitrijevic, G.~Djordjevic and M.~Milosevic,
Int. J. Mod. Phys. A \textbf{32}, no.05, 1750039 (2017)
doi:10.1142/S0217751X17500397
[arXiv:1607.04524 [gr-qc]].

\bibitem{Bilic:2018uqx}
N.~Bilic, D.~D.~Dimitrijevic, G.~S.~Djordjevic, M.~Milosevic and M.~Stojanovic,
JCAP \textbf{08} (2019), 034
doi:10.1088/1475-7516/2019/08/034
[arXiv:1809.07216 [gr-qc]].

\end{thebibliography}
\end{document}